\documentclass[conference]{IEEEtran}
\IEEEoverridecommandlockouts
\usepackage{booktabs} 
\usepackage{amssymb}
\usepackage{amsmath}
\usepackage[version=4]{mhchem}
\usepackage{graphicx}
\usepackage{multirow}
\usepackage{wrapfig}
\usepackage{color}
\usepackage{relsize}
\usepackage{colortbl}
\usepackage{bm}
\usepackage{soul}
\usepackage{mathrsfs}
\usepackage{subfigure}
\usepackage{enumerate}
\usepackage{paralist}
\usepackage{dsfont}
\usepackage[normalem]{ulem}
\usepackage{mathtools}
\usepackage{tabularx,booktabs,caption}
\usepackage[flushleft]{threeparttable}\usepackage{tabularx,booktabs,caption}
\usepackage[flushleft]{threeparttable}

\newcommand{\p}{\mathrm{p}}
\newcommand{\PP}{\mathrm{P}}
\newcommand{\E}{\mathrm{E}}
\newcommand{\e}{\mathrm{e}}

\newcommand{\bin}{\mathrm{Bin}}
\def\mathclap#1{\text{\hbox to 0pt{\hss$\mathsurround=0pt#1$\hss}}}
\makeatletter
\newcommand{\vastt}{\bBigg@{3}}
\newcommand{\vast}{\bBigg@{4}}
\newcommand{\Vast}{\bBigg@{5}}
\makeatother

\usepackage{cite}
\usepackage{amsmath,amssymb,amsfonts}
\usepackage{algorithmic}
\usepackage{graphicx}
\usepackage{textcomp}
\usepackage{xcolor}
\def\BibTeX{{\rm B\kern-.05em{\sc i\kern-.025em b}\kern-.08em
    T\kern-.1667em\lower.7ex\hbox{E}\kern-.125emX}}
\begin{document}

\title{Capacity Analysis of Molecular Communications with Ratio Shift Keying Modulation\\}

\author{\IEEEauthorblockN{M. Serkan Kopuzlu$^\ast$, M. Okan Araz$^\ast$, Ahmet R. Emirdagi$^\ast$, and Murat Kuscu}
\IEEEauthorblockA{Nano/Bio/Physical Information and Communications Laboratory (CALICO Lab) \\
	Department of Electrical and Electronics Engineering \\
	Koç University, Istanbul, Turkey \\
	\{mkopuzlu18, maraz18, aemirdagi18, mkuscu\}@ku.edu.tr}
\thanks{This work was supported in part by The Scientific and Technological Research Council of Turkey (TUBITAK) under Grant \#120E301, and European
	Union’s Horizon 2020 Research and Innovation Programme through the Marie Skłodowska-Curie Individual Fellowship under Grant Agreement \#101028935.}
\thanks{$^\ast$ These authors equally contributed to this work.}
}

\maketitle

\begin{abstract}
Molecular Communications (MC) is a bio-inspired communication technique that uses molecules to encode and transfer information. Many efforts have been focused on developing new modulation techniques for MC by exploiting distinguishable properties of molecules. In this paper, we investigate a particular modulation scheme where the information is encoded into the concentration ratio of two different types of molecules. To evaluate the performance of this so-called Ratio Shift Keying (RSK) modulation, we carry out an information theoretical analysis and derive the capacity of the end-to-end MC channel where the receiver performs ratio estimation based on ligand-receptor binding statistics in an optimal or suboptimal manner. The numerical results, obtained for varying similarity between the ligand types employed for ratio-encoding, and number of receptors, indicate that the RSK can outperform the concentration shift keying (CSK) modulation, the most common technique considered in literature, when the transmitter is power-limited. The results also indicate the potential advantages of RSK over other modulation methods under time-varying channel conditions, when the effects of the dynamic conditions are invariant to the type of the molecules. 
\end{abstract}


\section{Introduction}
Bio-inspired Molecular Communications (MC), which relies on biochemical molecules to encode and exchange information, is promising for interconnecting heterogeneous bio-nano things, e.g., engineered bacteria and nanobiosensors, thereby enabling unprecedented healtcare applications, such as intrabody continuous health monitoring within the Internet of Bio-Nano Things (IoBNT) framework \cite{malak2014communication,felicetti2016applications,akyildiz2015internet,akan2016fundamentals}. Over the last 15 years, there has been significant research interest in theoretical aspects of MC, such as channel modeling, detection and modulation techniques \cite{kuscu2019transmitter}. More recently, experimental studies have started to accompany this theoretical body of work \cite{kuscu2021fabrication, grebenstein2019molecular}. 

As the nature of information carriers in MC, i.e., molecules, is fundamentally different than that of the electromagnetic (EM) waves utilized in conventional communication technologies, researchers have developed novel modulation techniques that can exploit the distinguishable properties of molecules, such as concentration (concentration-shift-keying - CSK) \cite{kuran2011modulation}, molecule type (molecule-shift-keying - MoSK) \cite{kuran2012interference}, and release time of molecules from the transmitter (release-time-shift-keying - RTSK) \cite{murin2016communication}. Relatively less interest has been devoted to exploiting the concentration ratio between different types of molecules released simultaneously. This so-called ratio-shift-keying (RSK) modulation was proposed in \cite{kim2013novel}, where the authors considered encoding information into the concentration ratio of transmitted isomers that differ in the number of constituent monomers. However, to date, there has not been any study numerically evaluating the performance of the RSK modulation.

RSK can be speculated to have significant advantages over other modulation techniques at certain conditions of channel and transceivers. First, the same concentration ratio can be obtained with different absolute concentrations of individual types of molecules, implying more extended opportunities for molecule-efficient information exchange. Second, RSK can be more robust against dynamic variations in transmit power and channel impulse response (CIR), if the effect of these variations are molecule-type invariant. This can be exemplified by the mobile MC case when the diffusion coefficients of different types of molecules are equal. In that case, the time-varying CIR due to the mobility of the transceivers would be same for both types of molecules at all time instances, preserving the concentration ratio in the received signal. Likewise, RSK can be relatively robust in cases where the channel has specific enzymes that degrade both types of molecules at the same rate, which would not change the received concentration ratio. Similarly, we can consider cases where the transmitter, having a finite reservoir of molecules, or fluctuating molecule generation or harvesting mechanisms, can manifest time-varying transmission profiles in terms of absolute number of molecules transmitted. If the transmitter is able to maintain the transmitted concentration ratios at such conditions, the RSK can preserve its reliability. 

All the aforementioned advantages of RSK, however, are contingent upon the capacity of the receiver to detect the transmitted concentration ratios with high accuracy. In this paper, we investigate the performance of RSK modulation considering a physically-relevant receiver architecture, which is equipped with \textbf{single type of ligand receptors} interacting with different types of molecules, i.e., ligands, in a cross-reactive manner. Exploiting the difference in the affinities of the different types of ligands with the receptors, which is reflected to a difference in receptor-ligand bound time duration statistics, the receiver is able to estimate the received concentration ratio in a maximum-likelihood (ML) manner \cite{kuscu2019channel}. Due to the complexity concerns associated with this optimal ML estimation scheme, we also consider a suboptimal estimation method for the receiver, which is based on the biological kinetic proof-reading (KPR) mechanism \cite{kuscu2021detection}. 

To evaluate the performance of the RSK modulation, we carry out an information theoretical analysis and analytically derive the capacity of an end-to-end MC channel with a receiver employing optimal or suboptimal ratio estimation method. The numerical results obtained for a varying system settings, e.g., similarity between ligand types utilized for ratio-encoding, and number of receptors, are compared to the capacity of the MC channel employing a relatively more conventional CSK modulation. The results show that end-to-end MC channel with RSK manifests similar capacity as CSK, however, significantly outperforms CSK in cases where the transmit power (i.e., maximum concentration that the transmitter is able to send) is limited. These results indicate the potential of RSK in time-varying channel and transceiver conditions, and hint at its advantages over CSK and potentially other modulation techniques for the design of molecule-efficient MC systems. The performance of the suboptimal estimator which is revealed to be very close to that of the optimal scheme also indicates that the advantages of the RSK modulation can be realized with low computational complexity via biologically-relevant mechanisms.  

\section{Statistics of Ligand-Receptor Binding Reactions}
\label{sec:statistics}
Ligand-receptor interactions are key to communication and sensing in nature, as most biological cells, e.g., most bacteria, T-cells, express surface receptors as selective biorecognition elements, which undergo reversible reactions with specific types of molecules \cite{bialek2012biophysics}. These interactions are then translated into representations inside the cell, which in turn, inform the cell's subsequent actions. On the other hand, MC literature has so far mostly focused on receiver architectures that neglect the presence of receptors and ligand-receptor interactions. However, recent studies highlighted the significant impact of these interactions on the overall MC performance, and hinted at unique opportunities that can be obtained from their statistics \cite{kuscu2021fabrication, kuscu2016physical}. 

Ligand-receptor binding interactions can be formulated by a two-state continuous time Markov process with the states corresponding to the bound (B) and unbound (U) states of the receptor: 
\begin{equation}
	\ce{U  <=>[{c_L(t) k^+}][{k^-}] B},
	\label{equilibrium}
\end{equation}
where, $c_L(t)$ is the time-varying ligand concentration in the vicinity of the receptor, $k^+$ and $k^-$ are the binding and unbinding rates of the ligand-receptor pair, respectively. The dwell time of the receptor in states U and B is exponentially distributed with the rates $c_L(t) k^+$ and $k^-$, respectively. 

Due to the low-pass characteristics of the diffusion-based MC channel, the bandwidth of $c_L(t)$ is typically significantly smaller than the characteristic frequency of the binding reactions, i.e., $f_B=c_L(t ) k^+ +k^- $. Hence, ligand-receptor reactions can be assumed to be in equilibrium with a stationary ligand concentration in a time window of interest, and $c_L(t)$ can be simplified to $c_L$. In equilibrium conditions, the probability of a receptor being in the bound state is given as follows:
\begin{equation}
	\label{eq:success_prob_csk}
	\p_B=\frac{c_L}{c_L + K_D },
\end{equation}
where $K_D =k^-/k^+$ is the dissociation constant, which is inversely proportional to the ligand-receptor binding affinity. Considering that there are multiple receptors that do not interact with each other, and are exposed to the same ligand concentration, the number of bound receptors can be expressed as a binomial distribution, $N_B \sim \bin(\p_B,N_R)$. 


In the case of two different types of ligands in the receptors' vicinity, both ligands can bind the same receptors, but with different affinities, i.e., different $K_D$, which is reflected to the bound state probability of the receptors as follows
\begin{equation}
	\p_B=\frac{c_1/K_{D,1} + c_2/K_{D,2}}{1 + c_1/K_{D,1} + c_2/K_{D,2} },
	\label{eq:prob_binding_mixture}
\end{equation}
where $c_1$ and $c_2$ are the concentrations of different ligand types whose dissociation constants are denoted by $K_{D,1}$ and $K_{D,2}$ respectively. Due to the interchangeability of the summands, Eq. \eqref{eq:prob_binding_mixture} cannot be used to estimate the individual ligand concentrations, $c_1$ and $c_2$. As a result, in cases where different ligand types coexist in the channel, necessary statistics regarding individual ligand concentrations can only be obtained by analyzing the continuous history of ligand binding and unbinding events over receptors.

In diffusion limited cases, the characteristic rate of diffusion is much smaller than the ligand-receptor binding reaction rates, which allows us to simplify the binding rates for circular receptors as $k^+ = 4 Da$, with $D$ and $a$ denoting the diffusion constant of molecules, and the effective receptor size, respectively \cite{mora2015physical}. Assuming that the size difference between different ligand types is negligible, the diffusion constant, which then only depends on the temperature and viscosity of the channel medium, can be assumed equal for all ligand types. After these assumptions, the probability of observing a particular bound time duration $\p\left(\tau_{b}\right)$ in a single receptor can be written as
\begin{align}\label{eq:prob_taub}  
	\p(\tau_{b})   =  \sum_{j=1}^2 \alpha_j k_j^- \e^{-k_j^- \tau_{b}}  = \alpha_1 k^-_1 e^{-k^-_1 \tau_b} + \alpha_2  k^-_2 e^{-k^-_2 \tau_b}.
\end{align}
Here $ \alpha_1 = c_1 /(c_1+c_2)$ is the concentration ratio of the first ligand type, and $ \alpha_2 = 1-\alpha_1$ is the concentration ratio of the second ligand type. Then, the log-likelihood function for observing a set of bound time durations over $N_R$ independent receptors can be written as
\begin{equation}
	\mathscr{L}(\{\tau_b\}|\alpha) = \sum_{i=1}^{N_R} \ln \p(\tau_{b,i}),
	\label{eq:likelihood_boundtimedurations}
\end{equation}
where $\tau_{b,i}$ is the bound time duration observed on the $i^\text{th}$ receptor. 

\section{Channel Capacity}
\label{sec:channel_capacity}
Channel capacity is the maximum rate of information transfer in a communication channel and equal to the mutual information maximized over all input distributions. The input distribution that achieves the channel capacity is called the optimal input distribution, and denoted by $\PP^*(x)$. 

Under regularity conditions which are discussed in detail in \cite{clarke1994jeffreys,walker1969asymptotic}, $\PP_{N_S}^*(x)$ converges to Jeffreys Prior, $\PP_{jp}^*(x)$, as the number of independent copies of the signaling system, $N_S$, goes to infinity \cite{bernardo1979reference,clarke1994jeffreys,berger2009formal}. It was also shown that $\PP_{jp}^*(x)$ is proportional to the square root of the determinant of the Fisher information matrix \cite{jeffreys1946invariant}, indicating a direct link between information and estimation theory. By combining these two results, for one dimensional inputs, optimal input distribution asymptotically becomes proportional to the square root of the scalar Fisher information \cite{jetka2018information,komorowski2019limited}:
\begin{equation}
	{\PP}^{*}(x) \propto \sqrt{
		I(x)},
	\label{eq:opt_prob_dist}        
\end{equation}
resulting in the approximate channel capacity as follows
\begin{equation}
	\label{eq:asmyptotic_capacity_primitive}
	C_A^* = \log_2\bigg((2 \pi e)^{-\frac{1}{2}} \int_\mathcal{X} {\sqrt{ I(x)} dx} \bigg), 
\end{equation}
where $\mathcal{X}$ is the one-dimensional input symbol space.

\section{RSK Capacity}
\label{sec:rsk}
Using the formulation of the approximate channel capacity introduced in Section \ref{sec:channel_capacity}, here we derive the capacity of an MC channel where the transmitter employs RSK modulation by encoding information into the concentration ratio of two distinct types of ligands that it transmits into the diffusion-based MC channel, and the receiver estimates the concentration ratio from the bound time statistics of the resulting ligand-receptor interactions on its surface. 

Here we introduce a similarity parameter, $\gamma$, which is defined as the ratio of the unbinding rates of the distinct ligand types (type-1 and type-2), i.e., $\gamma = k_1^-/k_2^-$. Accordingly, the information is encoded into the concentration ratio of the first ligand type, i.e., $\alpha \in [0,1]$, where the subscript is omitted for ease of notation. 

The ML estimation of the concentration ratio of the incoming ligands from the receptor bound time intervals can be realized optimally. However, as shown in \cite{kuscu2019channel}, the complexity of optimal ML estimation may hamper its use in resource-constrained bio-nano machines. Therefore, we also calculate the channel capacity considering a receiver that employs a more practical suboptimal ratio estimator, which was introduced in \cite{kuscu2019channel}. 

\subsection{RSK Capacity with Optimal Ratio Estimation}
The ratio of the received ligand concentrations can be estimated in an ML manner by maximizing the likelihood of observing a set of bound time intervals $\{\tau_b\}$ over the input space, which, in this case, corresponds to the concentration ratio of ligands, i.e., $\alpha \in [0, 1]$. The log-likelihood of observing $\{\tau_b\}$ given the concentration ratio of the type-1 ligand can be written by transforming \eqref{eq:likelihood_boundtimedurations} as follows 
\begin{equation}
	\mathscr{L}(\{\tau_b\}|\alpha) = \sum_{i=1}^{N_R} \ln \biggl( k_2^- e^{- k_2^- \tau_{b,i}} \left(1 - \alpha + \alpha \gamma e^{(1 - \gamma) k_2^- \tau_{b,i})} \right) \biggr).    
\end{equation}

The Fisher information can then be derived from the log-likelihood function as follows
\begin{align}\label{fisher}
	I_{RSK}(\alpha) &= -\E \bigg[ \frac{\partial^2}{\partial \alpha^2} \mathcal{L} \left(\{\tau_b \} | \alpha\right) \bigg] \\ \nonumber
	&= N_R k_2^- \int_{0}^{\infty} \frac{\left(-1 + \gamma e^{(1 - \gamma) k_2^- \tau_{b} }\right)^2}{1 - \alpha + \alpha \gamma e^{(1 - \gamma) k_2^- \tau_{b}} } e^{-k_2^- \tau_b} d \tau_b.
\end{align}
By plugging this expression into Eqs. \eqref{eq:opt_prob_dist} and \eqref{eq:asmyptotic_capacity_primitive}, the optimal input distribution and the approximate channel capacity $C_{RSK}$ can be obtained, respectively, as follows
\begin{equation}
	{\PP}^{*}_{RSK}(\alpha) \propto \sqrt{
		I_{RSK}(\alpha)},
	\label{eq:opt_prob_dist_RSK}   
\end{equation}
\begin{equation} 
	C_{RSK} = \log_2 \bigg((2 \pi e)^{-\frac{1}{2}} \int_{0}^{1} {\sqrt{ I_{RSK}(\alpha)} d\alpha}\bigg).
\end{equation}

\subsection{RSK Capacity with Suboptimal Ratio Estimation}
Due to its high complexity, the optimal estimation of concentration ratios may not be feasible for resource-limited bio-nanomachines. Therefore, an alternative suboptimal estimation scheme for concentration ratios was recently proposed in \cite{kuscu2019channel}, whose estimation accuracy proved very close to the one of the optimal ML estimator. 

The low-complexity ratio estimator is based on a two-state KPR mechanism, which separates the long binding events, that are more likely to be generated by cognate (i.e., high-affinity) ligands, from the short-binding events. Accordingly, the estimator relies on the number of binding events that fall into either one of the short- or long-binding regimes, which are separated by a pre-defined time threshold $T$. This obviates the need for sampling the exact bound time durations and significantly reduces the complexity of the estimation.   

Given the concentration ratio of type-1 ligand, $\alpha$, the probability of observing a binding event with a bound time duration longer than $T$ can be written as 
\begin{align}\label{p_t_alpha}
	\p_T \equiv \p(\tau_b \geq{} T | \alpha) &= \alpha e^{-k_1T} + (1-\alpha) e^{-k_2T} \\ \nonumber 
	&= e^{-k_2T}\left(\alpha e^{(1-\gamma)k_2T} + 1- \alpha\right) 
\end{align}

The time threshold is a design parameter for the estimator, and can be tuned based on the unbinding rate of the low-affinity ligand, i.e. $T = \nu/k_1^-$, where $\nu$ is a proportionality constant. In \cite{kuscu2019channel}, the authors suggested that the optimal value of this constant is $\nu \approx 3$. Therefore, in the rest of the paper, we assume $T = 3/k_1^-$. Assuming that only a single binding event is sampled from each independent receptor, the number of binding events with $\tau_b > T$ is given by a binomial distribution
\begin{align}\label{binomial_suboptimal}	
	n_T \sim \mathcal{B}(\p_T,N_R).
\end{align}
Fisher information for the suboptimal concentration ratio estimator can then be written as follows
\begin{align}\label{fisher}
	&I_{RSK,sub}(\alpha) = -\E \bigg[ \frac{\partial^2}{\partial \alpha^2} \mathcal{L} \left(n_T  | \alpha\right) \bigg] = \left(e^{-\gamma k_2 T} - e^{-k_2 T}\right)^2 \\ \nonumber
	& \times \sum_{n_T=0}^{N_R}  \bigg[\frac{n_T}{\p_T^2} + \frac{N_R-n_T}{(1-\p_T)^2}  \bigg]  \binom{N_R}{n_T} \p_T^{n_T} \left(1-\p_T\right)^{N_R-n_T},
\end{align}
where the subscript $sub$ indicates the suboptimality of the estimation scheme employed by the receiver. 

Finally, $I_{RSK,sub}(\alpha)$ can be plugged into Eqs. \eqref{eq:opt_prob_dist} and \eqref{eq:asmyptotic_capacity_primitive} to obtain the optimal input distribution and the approximate capacity of the channel with the suboptimal ratio estimator as follows
\begin{equation}
	{\PP}^{*}_{RSK,sub}(\alpha) \propto \sqrt{
		I_{RSK,sub}(\alpha)},
	\label{eq:opt_prob_dist_RSK_sub}   
\end{equation}
\begin{equation} 
	C_{RSK,sub} = \log_2 \bigg({(2 \pi e)^{-\frac{1}{2} } \int_{0}^{1} {\sqrt{ I_{RSK,sub}(\alpha)} d\alpha}}\bigg).
\end{equation}

\section{CSK Capacity}
\label{sec:CSK_capacity}
In this section, we will investigate the approximate capacity of an MC channel with CSK modulation as a benchmark for in-depth evaluation of the RSK performance. In CSK, information is encoded into the concentration of a particular type of ligands, and the detection is performed through the sampling of the number of bound receptors in each signaling interval at a pre-defined sampling time. The sampling time is typically taken as the peak time of the ligand concentration in the vicinity of the receiver. Assuming equilibrium conditions for the ligand-receptor binding reaction at the sampling time, the number of bound receptors can be represented by a binomial distribution, i.e.,
\begin{equation}
	n_B \sim \mathcal{B}(\p_B,N),
\end{equation}
where, $n_B$ denotes the number of bound receptors, and $\p_B$ is the probability of a receptor to be in the bound state, as given by \eqref{eq:success_prob_csk}. The Fisher information, in this case, can be calculated as follows
\begin{align}
	\label{eq:fisher_CSK}
	&I_{CSK}(c) = -\E \bigg[ \frac{\partial^2}{\partial c^2} \mathcal{L} \left(n_B | c \right) \bigg] \\ \nonumber
	&= -\sum_{n_B=0}^{N_R} \Bigg( \frac{\partial \p_B^2}{\partial c^2}  \bigg[\frac{n_B}{\p_B} - \frac{N_R-n_B}{1-\p_B} \bigg] \\ \nonumber
	&- \bigg(\frac{\partial \p_B}{\partial c} \bigg)^2 \bigg[\frac{n_B}{\p_B^2} + \frac{N_R-n_B}{(1-\p_B)^2} \bigg] \Bigg) \times \binom{N_R}{n_B} \p_B^{n_B} (1-\p_B)^{N_R-n_B}.
\end{align}

\begin{figure*}[h]
	\centering
	\subfigure[]{\label{fig:capacity_gamma}\includegraphics[width=0.32\linewidth]{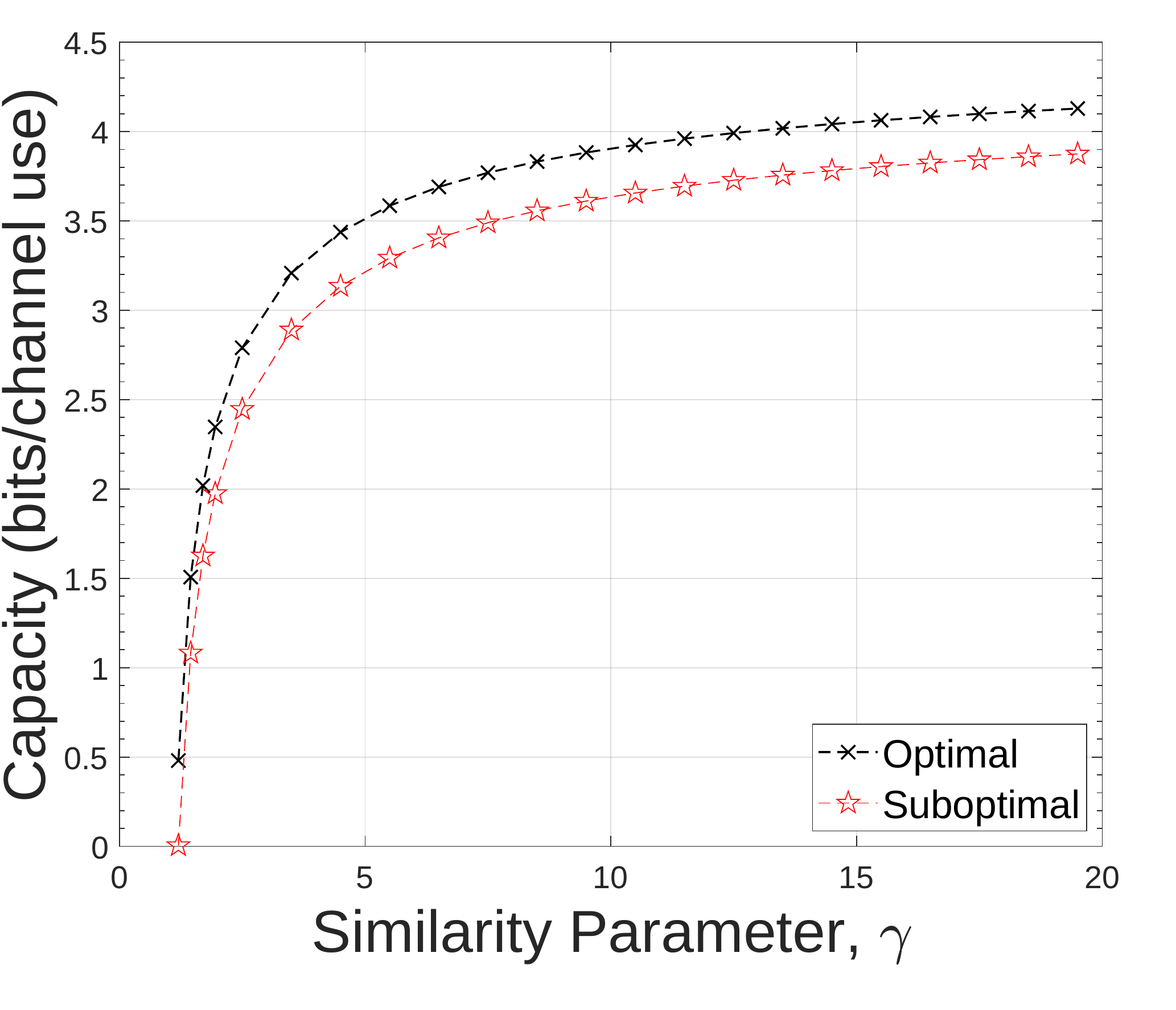}}\quad
	\subfigure[]{\label{fig:capacity_N}\includegraphics[width=0.32\linewidth]{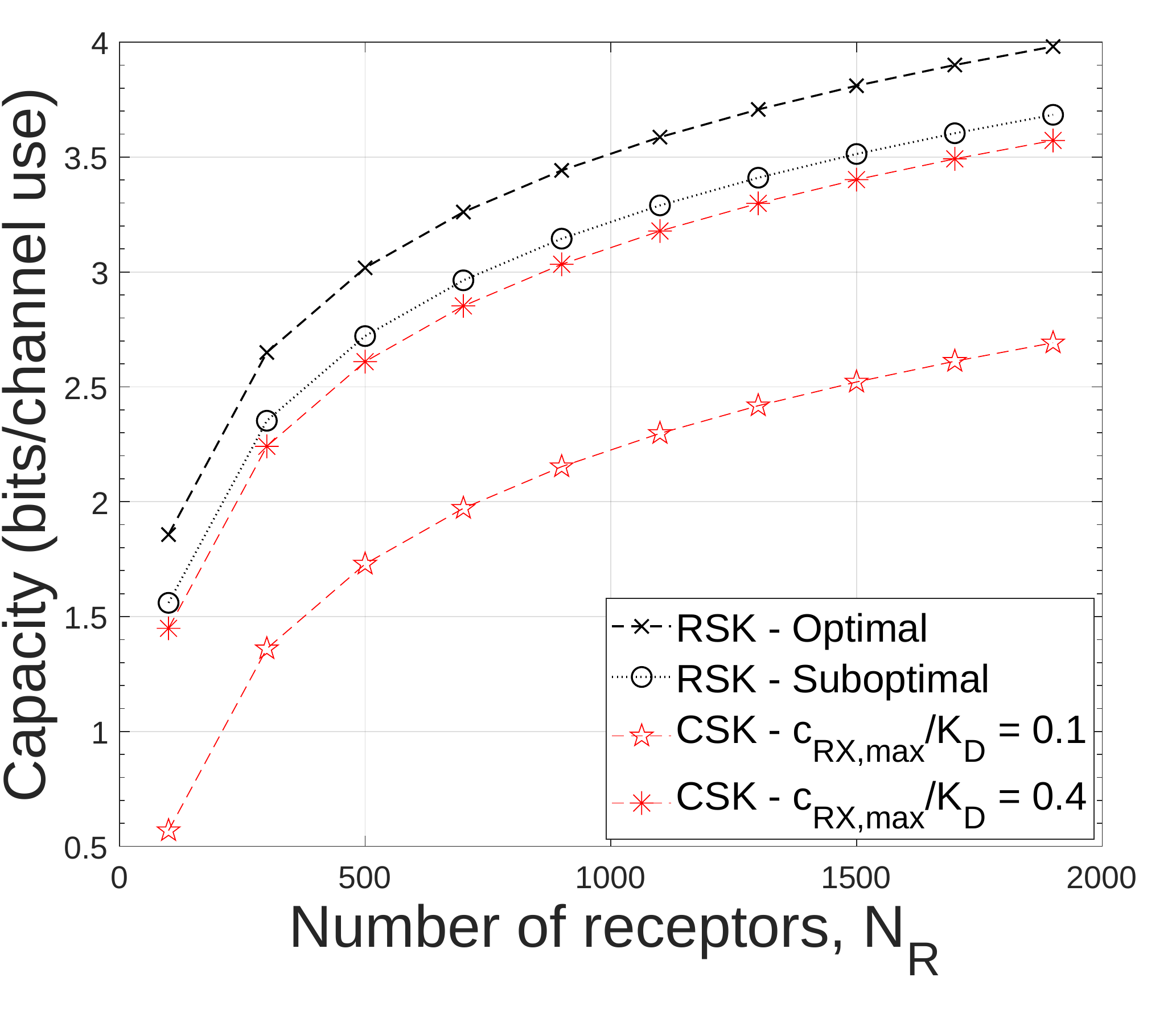}}\quad
	\subfigure[]{\label{fig:capacity_CSK}\includegraphics[width=0.32\linewidth]{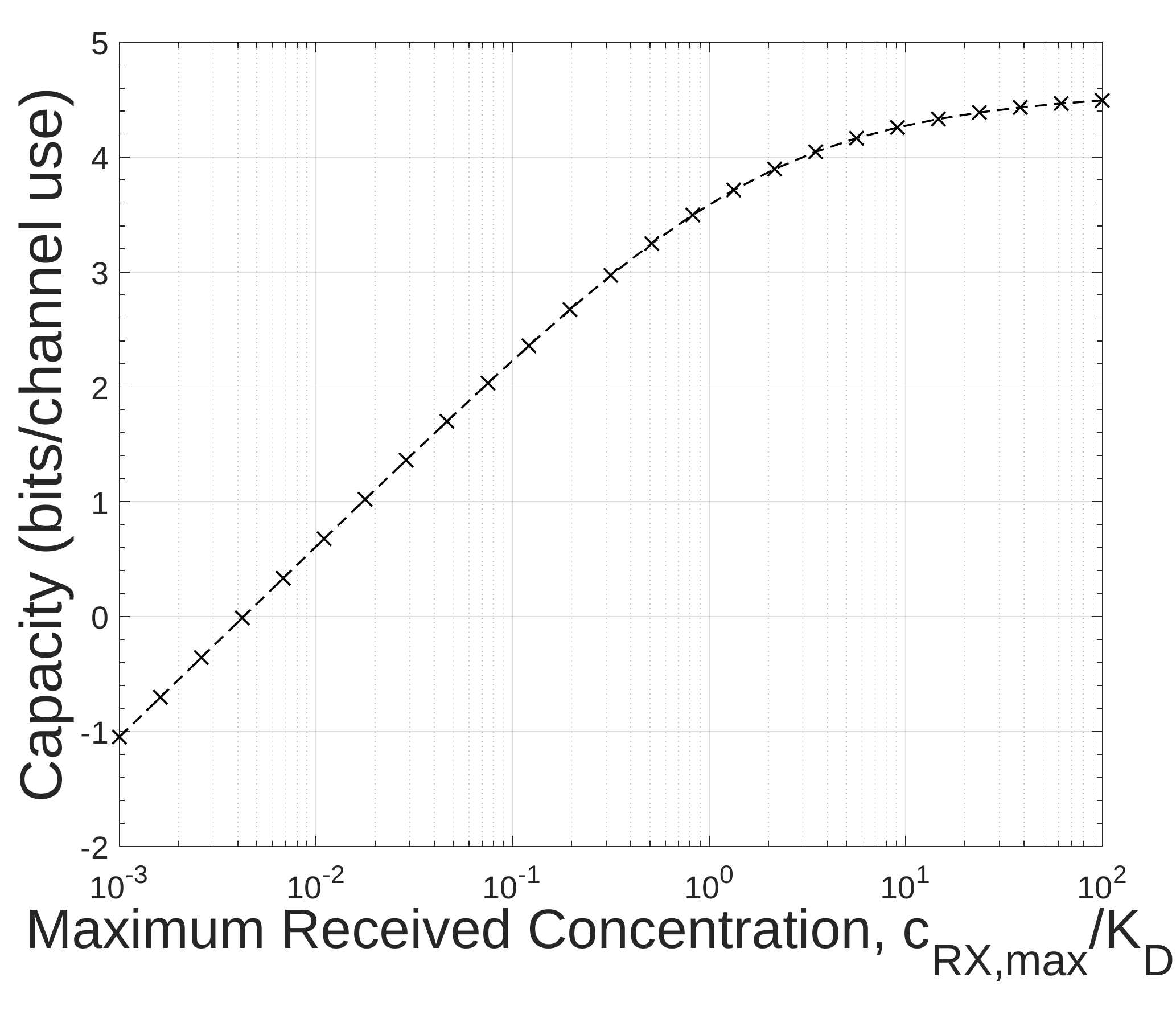}}
	
	\caption{Channel capacity (a) with RSK as function of similarity parameter, $\gamma$, (b) with RSK and CSK as function of number of receptors, $N_R$, (c) with  CSK as function of maximum received concentration $c_{RX,max}$.}
\end{figure*}

In an MC system with a power limited transmitter, input symbol space in equation \eqref{eq:asmyptotic_capacity_primitive} is limited by the maximum concentration of ligands that the transmitter can release to the channel. In this case, the approximate capacity of the channel can be obtained as follows
\begin{equation} 
	C_{CSK} = \log_2 \Bigg({(2 \pi e)^{-\frac{1}{2} } \int_{0}^{c_{RX,max}} {\sqrt{ I_{CSK}(c) } dc\Bigg)}},
\end{equation}
where $c_{RX,max}$ is the ligand concentration in the vicinity of the receptors at the sampling time, which corresponds to the maximum ligand concentration that can be transmitted by the transmitter, scaled by the CIR of the diffusion-based MC channel. Finally, optimal input distribution can be obtained by plugging Eq. \eqref{eq:fisher_CSK} in Eq. \eqref{eq:opt_prob_dist} as follows
\begin{equation}
	{\PP}^{*}_{CSK}(c) \propto \sqrt{
		I_{CSK}(c)}.
	\label{eq:opt_prob_dist_CSK}   
\end{equation}


\section{Numerical Results}
In this section, we numerically evaluate the channel capacity of MC with RSK and CSK under different system settings. RSK capacity is analyzed with respect to similarity parameter, and the number of receptors where CSK capacity is analyzed with respect to maximum received concentration and number of receptors. We also present the corresponding optimal input distributions of RSK and CSK. Default values of the number of receptors and the similarity parameter used in the analyses are $N_R = 1000$ and $\gamma = 5$, respectively.

\subsection{Channel Capacity}
We first analyze the impact of the \textbf{ligand similarity parameter}, $\gamma = k_1^-/k_2^-$, on the capacity of the MC channel with RSK. The results are provided in Fig. \ref{fig:capacity_gamma} for cases where the receiver employs either optimal or suboptimal estimator for concentration ratios. Increasing $\gamma$ corresponds to decreasing similarity between the ligand types in terms of their unbinding rates from the receptors, which in turn increases their distinguishability by the receiver. As is clear in Fig. \ref{fig:capacity_gamma}, decreasing similarity increases the accuracy of both the optimal and the suboptimal estimators, leading to higher channel capacities, which saturate around $4$ bits/channel use. We also note that when $\gamma \approx 1$, the different ligand types are hardly distinguishable from each other, and thus, the approximate channel capacity approaches $0$. However, even a slight difference in the unbinding characteristics of the ligands significantly improves the channel capacity. More importantly, the channel capacity obtained with the suboptimal estimator is very close to that obtained with the optimal estimator, although the former has much lower complexity than the latter, as detailed in Section \ref{sec:rsk}. This underlies the feasibility of RSK modulation for resource-constrained bio-nanomachines, as the suboptimal estimator can be implemented by a simple single-threshold KPR scheme, more complex versions of which are already utilized in natural cells \cite{mckeithan1995kinetic}. 
\begin{figure*}[t]
	\centering
	\subfigure[]{\label{fig:OID_RSK_optimal}\includegraphics[width=0.32\linewidth]{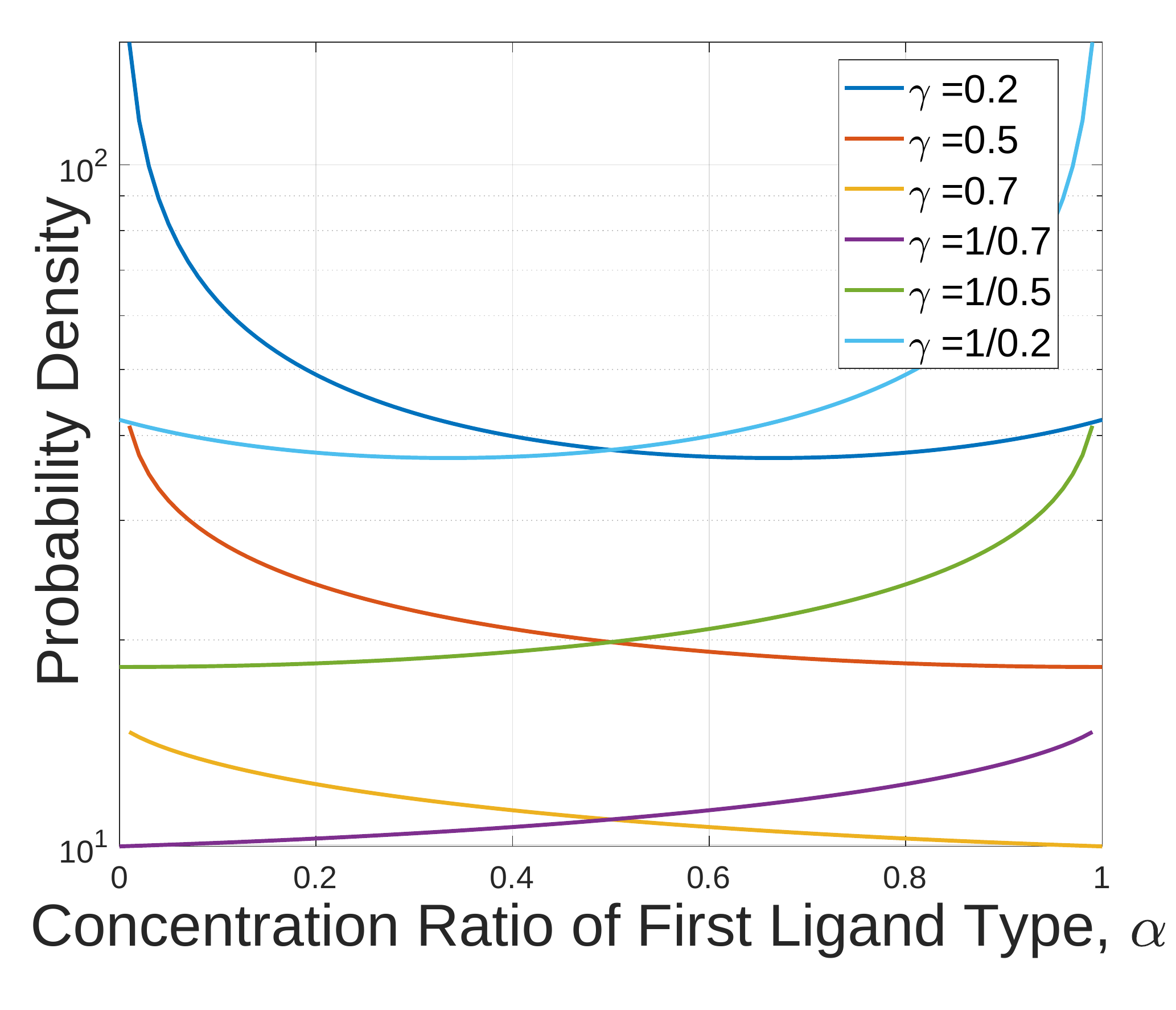}}\quad
	\subfigure[]{\label{fig:OID_RSK_suboptimal}\includegraphics[width=0.32\linewidth]{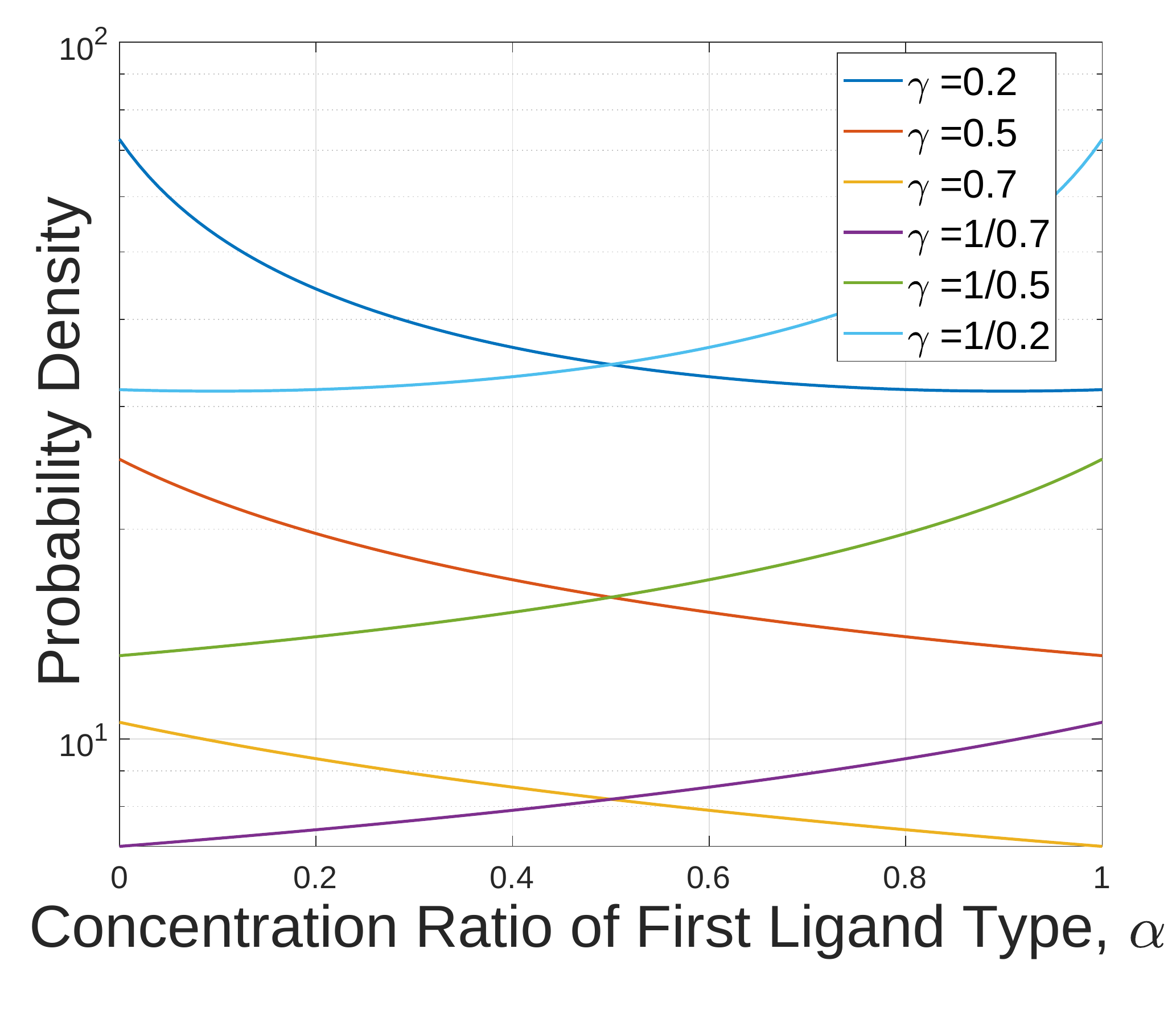}}\quad    
	\subfigure[]{\label{fig:OID_CSK}\includegraphics[width=0.32\linewidth]{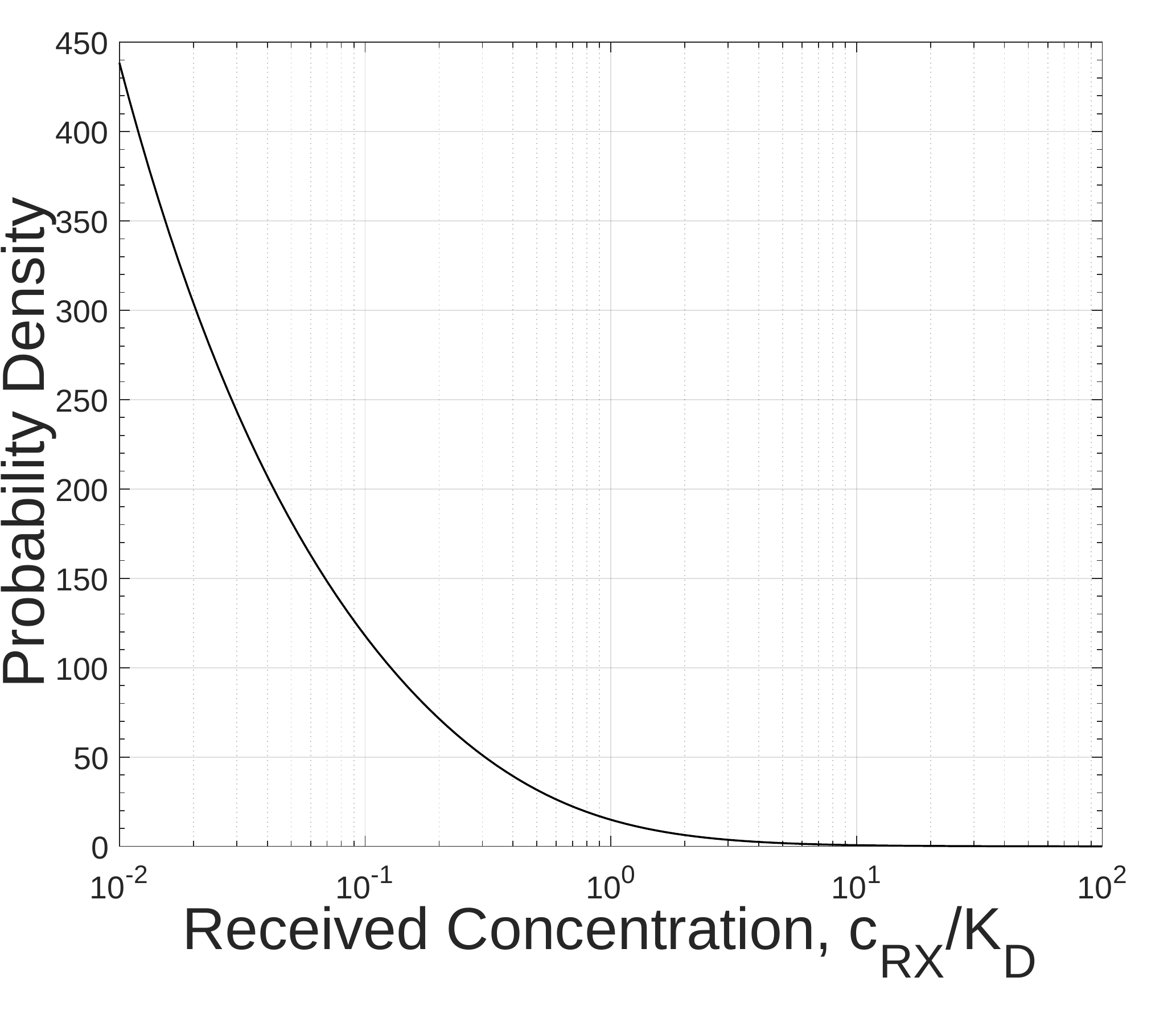}}
	\caption{Optimal input distribution for MC channel (a) with RSK using optimal estimator, (b) with RSK using suboptimal estimator, (c) with CSK.}
\end{figure*}

Our next analysis concerns the impact of the \textbf{number of receptors}, $N_R$, on the channel capacity. As the receptors are considered to be independent of each other, $N_R$ gives the \textbf{number of independent samples} taken from the receptors in each symbol interval for estimating the ratio in the case of RSK, and the concentration in the case of CSK. Accompanying this analysis, results of which are provided in Fig. \ref{fig:capacity_N}, is a comparison between RSK and CSK for cases when the transmitter is power-limited, i.e., the number of molecules it can transmit is upper-bounded. First, we observe that the number of independent samples, i.e., $N_R$, has a significant effect on the capacity, as taking more independent samples improves the accuracy of the estimation performed by the receiver. The power limitation of the transmitter is translated into an upper-bounded ligand concentration in the vicinity of receptors, as the free diffusion channel is deterministic in terms of molecule concentration. Accordingly the maximum received concentration is set to $c_{RX,max} = 0.1~K_D$ and $c_{RX,max} = 0.4~K_D$ for CSK modulation. We neglect the effect of power limitation on the RSK performance, as the concentration ratio is invariant to the total number of molecules released by the transmitter. The results are in accordance with our discussion in the Introduction, such that RSK becomes advantageous in terms of channel capacity when the received concentration is upper-bounded. This advantage can be crucial particularly when the transmitter has a limited molecule reservoir or relies on fluctuating molecule harvesting or production processes. Moreover, RSK's advantages over CSK can be prominent for mobile MC cases, where the mobility of the transmitter and/or the receiver results in a time-varying CIR and received concentration profile, deteriorating the MC performance.

For the completeness of the analysis, we also provide the approximate channel capacity for CSK as a function of maximum received concentration $c_{RX,max}$ in Fig. \ref{fig:capacity_CSK}. The results show that the asymptotic channel capacity obtained when the $c_{RX,max}$ is much larger than the dissociation constant of the ligand-receptor pair $K_D$ saturates around $4.5$ bits/channel use, which is only slightly higher than the asymptotic channel capacity of the RSK obtained when the similarity between the utilized ligand types is low (see Fig. \ref{fig:capacity_gamma}).

\subsection{Optimal Input Distribution} 
The optimal input distribution achieving the channel capacity for RSK are given in Figs. \ref{fig:OID_RSK_optimal} and \ref{fig:OID_RSK_suboptimal}, for cases where the receiver utilizes optimal and suboptimal estimators, respectively. Here, we take the input as the ratio of the first ligand type's concentration to the total ligand concentration, i.e., $\alpha \in [0,1]$, and the optimal input distributions are given for different values of the similarity parameter, $\gamma = k_1^-/k_2^-$. Our first observation is a straightforward one such that the optimal input distribution is symmetric for $\gamma = a$ and $\gamma = 1/a$, with $a \in (0,1)$. The second observation is that the optimal input distribution favors the ligand type which has lower affinity with the receptors, i.e., the one having higher unbinding rate. This can be explained through the fact that it is less likely to find a receptor bound to a ligand with higher unbinding rate at equilibrium, in turn, decreasing the number of samples informative of this particular ligand type. As a result, the optimal input distribution shifts the concentration ratio towards the less-likely binding ligand, leading to an increase in their proportion among all the bound ligands at equilibrium. Moreover, this preference becomes more salient as the similarity between the ligand types decreases. The same trend can be observed in both optimal and suboptimal cases, although the preference for the less cognate ligand is more prominent in the optimal case.

The optimal input distribution for MC channel with CSK is given in Fig. \ref{fig:OID_CSK}. Our numerical analysis for the calculation of optimal input distribution does not have a constraint on the maximum received concentration $c_{RX,max}$ where the input is the ligand concentration in the vicinity of the receiver. We observe that the optimal input distribution favors low ligand concentrations. This is consistent with the results of Einolghozati et al. in \cite{einolghozati2011capacity}.

\section{Conslusion}

We performed an information-theoretical analysis of the MC channel with RSK modulation considering two different ratio estimation schemes, varying in their optimality and complexity, for a ligand-receptor-based receiver. The performance was numerically compared to that of CSK modulation in terms of the corresponding channel capacity. The results demonstrated that RSK modulation outperforms CSK modulation particularly when the transmitter is power-limited, such that the received ligand concentration is upper-bounded. The future work will address the performance of RSK modulation at time-varying end-to-end channel conditions, when the effects of these dynamic conditions are invariant to the type of the utilized molecules, hence their concentration ratio.

\bibliographystyle{IEEEtran}
\bibliography{RSKpaper_arxiv} 

\end{document}